\documentclass[conference, 10pt]{IEEEtran}
\IEEEoverridecommandlockouts
\usepackage{cite}
\usepackage{amsmath,amssymb,amsfonts}
\usepackage{algpseudocode}
\usepackage{graphicx}
\usepackage{textcomp}
\usepackage{xcolor}
\usepackage{braket}
\usepackage{url}
\usepackage{hyperref}
\hypersetup{
  colorlinks   = true, 
  urlcolor     = blue, 
  linkcolor    = blue, 
  citecolor   = blue 
}

\def\BibTeX{{\rm B\kern-.05em{\sc i\kern-.025em b}\kern-.08em
    T\kern-.1667em\lower.7ex\hbox{E}\kern-.125emX}}
\begin{document}

\title{Quantum-Train with Tensor Network \\ Mapping Model and Distributed Circuit Ansatz
}


\author{
\IEEEauthorblockN{
     Chen-Yu Liu \IEEEauthorrefmark{1}\IEEEauthorrefmark{6}, 
    Chu-Hsuan Abraham Lin\IEEEauthorrefmark{3}\IEEEauthorrefmark{7},
    Kuan-Cheng Chen\IEEEauthorrefmark{3}\IEEEauthorrefmark{4}\IEEEauthorrefmark{8}
}

\IEEEauthorblockA{\IEEEauthorrefmark{1}Graduate Institute of Applied Physics, National Taiwan University, Taipei, Taiwan}
\IEEEauthorblockA{\IEEEauthorrefmark{3}Department of Electrical and Electronic Engineering, Imperial College London, London, UK}
\IEEEauthorblockA{\IEEEauthorrefmark{4}Centre for Quantum Engineering, Science and Technology (QuEST), Imperial College London, London, UK}

\IEEEauthorblockA{Email:\IEEEauthorrefmark{6} d10245003@g.ntu.edu.tw, \IEEEauthorrefmark{7} abraham.lin23@imperial.ac.uk, \IEEEauthorrefmark{8}kuan-cheng.chen17@imperial.ac.uk}

}

\maketitle

\begin{abstract}
In the Quantum-Train (QT) framework, mapping quantum state measurements to classical neural network weights is a critical challenge that affects the scalability and efficiency of hybrid quantum-classical models. The traditional QT framework employs a multi-layer perceptron (MLP) for this task, but it struggles with scalability and interpretability. To address these issues, we propose replacing the MLP with a tensor network-based model and introducing a distributed circuit ansatz designed for large-scale quantum machine learning with multiple small quantum processing unit nodes. This approach enhances scalability, efficiently represents high-dimensional data, and maintains a compact model structure. Our enhanced QT framework retains the benefits of reduced parameter count and independence from quantum resources during inference. Experimental results on benchmark datasets demonstrate that the tensor network-based QT framework achieves competitive performance with improved efficiency and generalization, offering a practical solution for scalable hybrid quantum-classical machine learning.

\end{abstract}


\section{Introduction}

In recent years, Quantum Computing and Quantum Machine Learning (QML) have shown substantial potential in enhancing learning efficiency and flexibility by integrating diverse computational architectures \cite{biamonte2017quantum, dunjko2016quantum}. Leveraging the unique properties of quantum systems, such as superposition and entanglement, QML can perform parallel computations across multiple basis states simultaneously\cite{lau2017quantum}. This capability has enabled a broad spectrum of applications, including classification tasks \cite{qmlapp2}, reinforcement learning \cite{chen2020variational}, time-series forecasting \cite{lin2024quantum}, and the incorporation of quantum algorithms into various computational frameworks \cite{liu2023learning, liu2023practical, liu2023reinforcement}.

In conventional QML approaches, data is typically introduced into quantum circuits through encoding techniques like gate-angle encoding and amplitude encoding \cite{huang2021power}. However, encoding large datasets into quantum circuits remains a significant challenge due to the limited number of available qubits and the constraints on circuit depth imposed by the coherence times of current quantum systems. Beyond data encoding limitations, once a QML model is trained, the inference phase often necessitates the use of quantum hardware (commonly accessed via cloud-based platforms), where hybrid computations occur layer-by-layer. This dependency can result in inefficiencies, particularly for time-sensitive applications such as real-time decision-making in autonomous vehicles.

To address these challenges, the Quantum-Train (QT) framework has been proposed as an innovative “learning-wise” hybrid quantum-classical architecture. The central concept is to decouple the quantum processing from direct data interaction by utilizing a quantum neural network (QNN) to generate the weights of a classical machine learning (ML) model. This approach effectively circumvents data encoding issues, as the data is processed entirely within the classical model, bypassing the need for direct quantum data input. Moreover, it eliminates the requirement for quantum hardware during the inference stage, making the final model fully classical and independent of quantum resources post-training. These characteristics position QT as a highly practical and efficient framework for near-term QML applications \cite{liu2024training, liu2024quantum, liu2024qtrl, lin2024quantum}.

\begin{figure}[!t]
\centering
\includegraphics[scale=0.55]{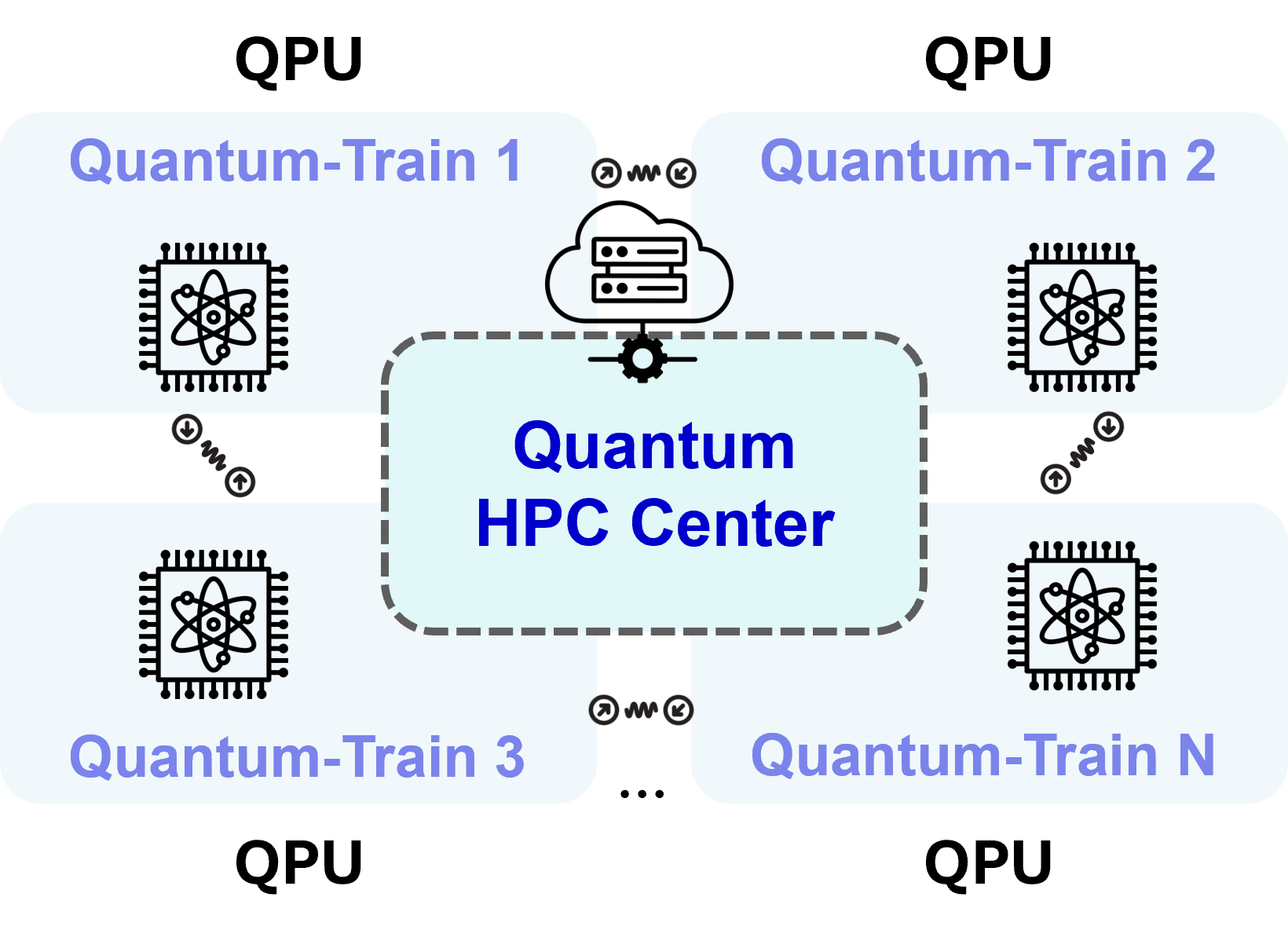}
\caption{Conceptual diagram of the distributed Quantum-Train architecture applied in Quantum High-Performance Computing (HPC) for large-scale quantum machine learning problems.
 }
\label{fig:concept}
\end{figure}



A critical step in the QT framework is mapping the measured probabilities of quantum basis states to the weights of the target machine learning model. In previous implementations, this mapping has primarily been achieved using a multi-layer perceptron (MLP) architecture. However, due to the high-dimensional nature of the quantum states generated by the QNN, MLPs may struggle to efficiently capture the complex structure of these probabilities. To address this, Tensor Networks (TN) present a compelling alternative due to their ability to efficiently represent weakly interacting quantum states and their widespread use in quantum circuit simulations\cite{pan2022simulation,chen2024cutn}.

Another significant challenge in QNNs is the extensive qubit requirement, which limits scalability for both simulations and real quantum hardware. To address these limitations, a distributed circuit ansatz can be employed \cite{ferrari2023modular, burt2024generalised}. Additionally, the possible integration of distributed quantum computing with hybrid quantum HPC architectures has been explored in recent work \cite{chen2024qcq}. As illustrated in Fig. \ref{fig:concept}, the concept of large-scale distributed Quantum-Train (QT) for a Quantum HPC center aims to tackle large-scale machine learning problems. This approach involves partitioning the quantum circuit into smaller sub-circuits, thereby reducing overall qubit usage and allowing the results to be efficiently combined through tensor products after computation.

In this work, we propose two key enhancements to the QT framework: (1) substituting the MLP mapping model with a Tensor Network structure to improve expressive power while reducing the number of parameters, and (2) designing the QNN ansatz in a distributed manner, which not only accelerates simulations but also decreases qubit requirements for real quantum hardware applications.

\section{Quantum-Train}
\label{sec:qt}

\begin{figure*}[ht]
\centering
\includegraphics[scale=0.31]{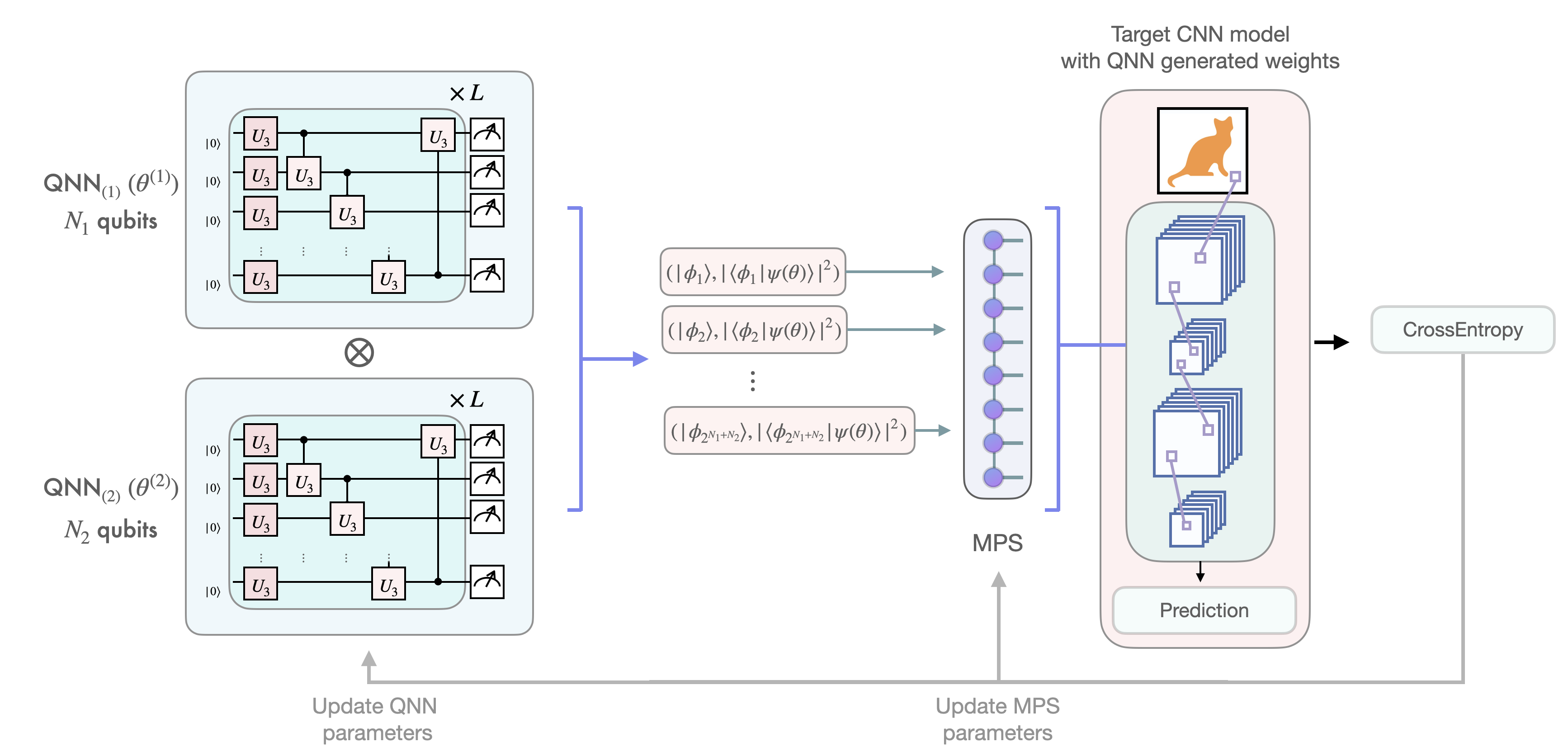}
\caption{The scheme of the QT framework with a TN mapping model and a distributed circuit ansatz is illustrated as follows. The process begins with two distributed quantum circuits, whose results are combined via a tensor product. The basis information and the corresponding measurement probabilities are then input into the MPS structure. The MPS subsequently generates the weights for the target CNN model.
}
\label{fig:scheme}
\end{figure*}

The traditional QT framework is outlined as follows: Consider a classical neural network (NN) with parameters $\vec{\omega} \in \mathbb{R}^{m}$, defined as
\begin{equation}
\vec{\omega} = (\omega_1, \omega_2, \ldots, \omega_m).
\end{equation}
A QNN with $N = \lceil \log_2 m \rceil$ qubits is then constructed using a real amplitude ansatz, expressed as:
\begin{equation}
\label{eq:ansatz}
|\psi(\theta) \rangle = \left(\prod_i \text{CNOT}^{i, i+1} \prod_j R_y^j \right)^L |0\rangle^{\otimes N}.
\end{equation}
Rather than updating all $m$ parameters, as is done in conventional ML, the QT approach utilizes the quantum state $|\psi (\theta) \rangle$ to produce $2^N$ distinct measurement probabilities $|\langle \phi_i | \psi (\theta) \rangle|^2$, where $i \in \{1, 2, \ldots, 2^N\}$, and $|\phi_i \rangle$ represents the $i$-th basis state. These probabilities are passed through a mapping model $G_{\beta}$, a classical MLP network with parameters $\beta$.

The first $m$ measurement probabilities, along with their corresponding basis states $|\phi_i \rangle$, are transformed from values in the interval $[0, 1]$ to the range $(-\infty, \infty)$ using the following relation:
\begin{equation}
G_{\beta} (|\phi_i \rangle, |\langle \phi_i | \psi (\theta) \rangle|^2) = \omega_i, \quad i = 1, 2, \ldots, m.
\end{equation}
This shows that the parameters $\omega_i$ of the target model are derived from the QNN state $|\psi(\theta) \rangle$ and the mapping model $G_{\beta}$. Crucially, the number of tunable parameters in both $\theta$ and $\beta$ scales as $O(\text{polylog}(m))$ \cite{liu2024quantum}, enabling efficient training of the target model with only $O(\text{polylog}(m))$ parameters, rather than updating all $m$ parameters directly.

\subsection{Tensor Network Mapping Model}

TNs have achieved significant success in quantum many-body physics, machine learning tasks, and even quantum circuit simulations. Due to their efficiency in representing high-dimensional data, TNs are expected to more effectively represent the mapping between quantum measurement probabilities and classical NN weights. Following the formulation in \cite{stoudenmire2016supervised}, the matrix product state (MPS) decomposition of the mapping model’s weight tensor $W$ is expressed as:
\begin{equation}
W_{s_1, s_2, \hdots, s_{N+1}} = \sum_{{ \alpha}} A^{\alpha_1}_{s_1} A^{\alpha_1 \alpha_2}_{s_2} \hdots A^{\alpha_{N}}_{s_{N+1}},
\end{equation}
where $N$ corresponds to the number of qubits, as mentioned in the previous section, since the vector representation of $(|\phi_i \rangle, |\langle \phi_i | \psi (\theta) \rangle|^2)$ has the shape $N+1$. With the feature map
\begin{equation}
\Xi^{s_1, s_2, \hdots, s_{N+1}}(\mathbf{x}) = \xi^{s_1}(x_1) \otimes \xi^{s_2}(x_2) \otimes \hdots \xi^{s_{N+1}}(x_{N+1}),
\end{equation}
and
\begin{equation}
\xi^{s_j}(x_j) = \begin{bmatrix}
x_j \\
1-x_j
\end{bmatrix},
\end{equation}
the mapping model is now given by
\begin{equation}
G(|\phi \rangle, |\langle \phi | \psi (\theta) \rangle|^2) = W \cdot \Xi(|\phi \rangle, |\langle \phi | \psi (\theta) \rangle|^2) = \vec{\omega}.
\end{equation}
In this formulation, the MPS tensors $A^{\alpha_j}_{s}$ serve as the tunable parameters of the TN mapping model, where each virtual index $\alpha_j$ is associated with a bond dimension $r$.

\subsection{Distributed Circuit Ansatz}

The number of qubits required for a QNN can sometimes become too large for efficient evaluation, both in classical simulations and on quantum hardware, depending on the specific task. For instance, the CIFAR-10 classification task within the QT proposal in \cite{liu2024quantum} requires $\lceil \log_2 285226 \rceil = 19$ qubits. Although 19 qubits is a feasible size, evaluating a quantum circuit of this scale is significantly slower than using circuits with 9 or 10 qubits.

Since the goal of the QNN in this case is to generate $2^{19}$ measurement probabilities, the circuit can be decomposed into two sub-circuits with 9 and 10 qubits, respectively, given that the Hilbert space dimension satisfies $2^{19} = 2^{10} \times 2^9$. This decomposition is expressed as:
\begin{equation}
| \psi \rangle_{N = 19} = |\psi \rangle_{N=9} \otimes | \psi \rangle_{N=10},
\end{equation}
where the measurement probabilities for the full system are given by:
\begin{eqnarray}
&&|\langle s_1 s_2 \hdots s_{19} | \psi \rangle_{N=19}|^2 = \\
&&|\langle s_1 s_2 \hdots s_{9} | \psi \rangle_{N=9}|^2  |\langle s_{10} \hdots s_{19} | \psi \rangle_{N=10}|^2.
\end{eqnarray}

The distributed ansatz is then written as:
\begin{eqnarray}
|\psi(\theta) \rangle_{N} = |\psi(\theta^{(1)}) \rangle_{N_1} \otimes |\psi(\theta^{(2)}) \rangle_{N_2},
\end{eqnarray}
where $|\psi(\theta^{(1)}) \rangle_{N_1}$ and $|\psi(\theta^{(2)}) \rangle_{N_2}$ correspond to the components as defined in Eq.~\ref{eq:ansatz}. Following the training procedure outlined in the original QT framework, the QT method with a TN mapping model and distributed circuit ansatz is illustrated in Fig.~\ref{fig:scheme}.

\section{Numerical Results and Discussion}
\label{sec:nrd}

\begin{figure}[ht]
\centering
\includegraphics[scale=0.38]{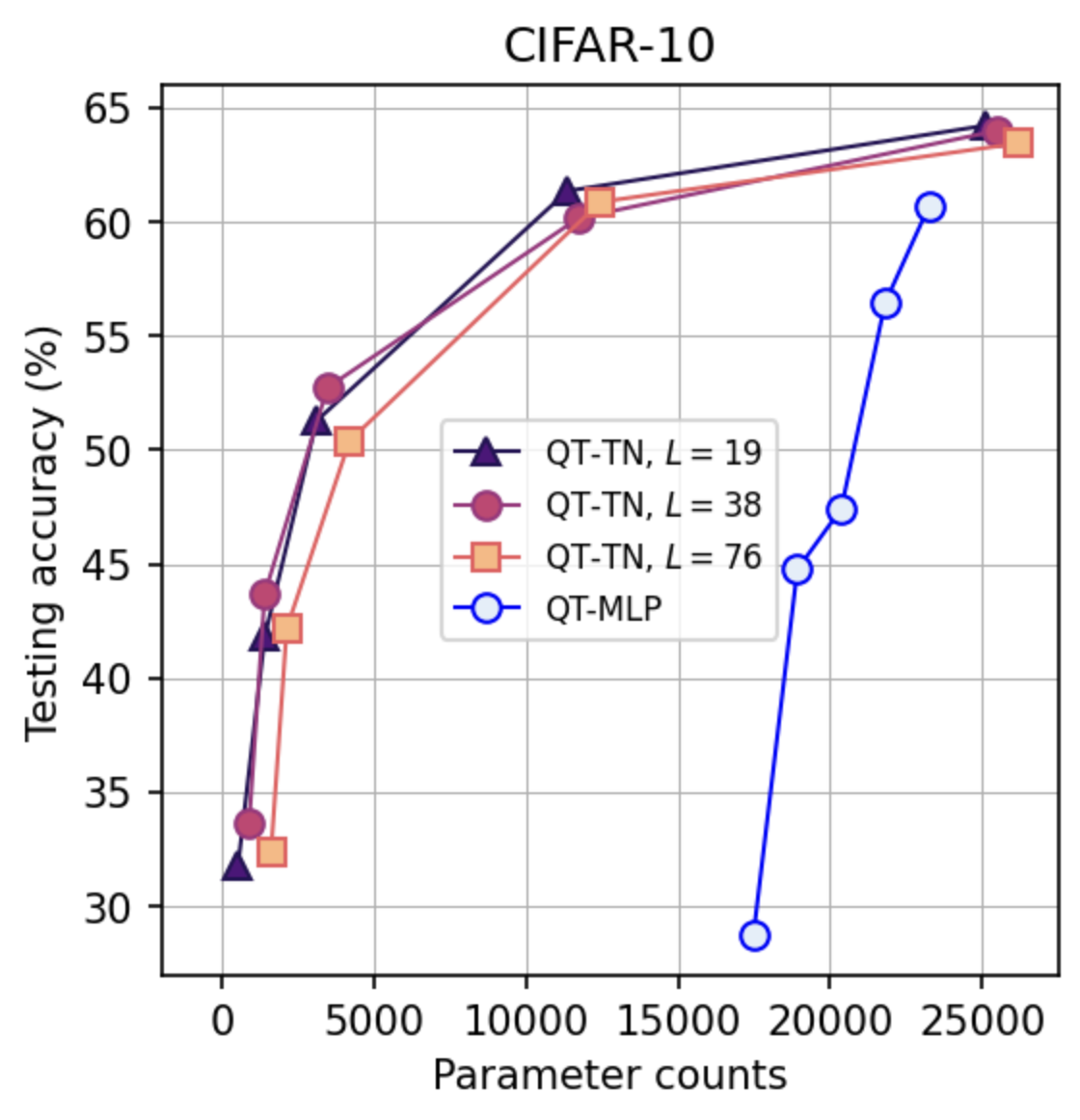}
\caption{The comparison between the TN mapping model with the distributed circuit ansatz and the MLP mapping model with the conventional circuit ansatz is presented. The blue line represents the results obtained in \cite{liu2024quantum}.
 }
\label{fig:result}
\end{figure}

The effectiveness of the TN mapping model and distributed circuit ansatz is evaluated using the following setup: The target model is the same convolutional neural network (CNN) from the original QT paper \cite{liu2024quantum}. For the CIFAR-10 classification task, the model contains 285226 parameters. The results for the QT framework with the MLP mapping model, shown as the blue line in Fig.~\ref{fig:result} (QT-MLP), are taken from \cite{liu2024quantum} as the baseline for comparison. The QNN repetition counts $L$, ranging from small to large parameter counts, are $\{19, 95, 171, 247, 323 \}$. In comparison, the QT-TN distributed ansatz results with $L \in \{19, 38, 76\}$ are displayed. The parameter variations are controlled by the bond dimensions of the MPS, with $r \in \{2, 4, 8, 16, 24\}$. It is evident that the QT-TN results outperform those of the MLP mapping model. At a testing accuracy of approximately $60\%$, the QT-TN model requires slightly more than 10000 parameters, whereas the MLP mapping model requires approximately 23000 parameters. Nevertheless, both approaches offer significant compression compared to the original parameter count of 285226.

An interesting property can also be observed in Fig.~\ref{fig:result}. When comparing the slope of the QT-TN results with respect to increasing parameter counts, the slope of the QT-MLP results is noticeably steeper. The QT-TN results correspond to increasing the “classical” TN parameters, specifically the bond dimension, while the QT-MLP results reflect an increase in “quantum” parameters, specifically the QNN block repetition $L$, as shown in \cite{liu2024quantum}. This indicates that the impact of increasing classical and quantum parameters is not equivalent: increasing quantum parameters results in a greater improvement in accuracy per unit of parameter increase. However, in the current noisy intermediate-scale quantum (NISQ) era, increasing the number of QNN layers requires a longer coherence time for qubits on real quantum hardware, as well as extended circuit simulation time on classical simulators. In this context, although the TN mapping model with an increased bond dimension yields less performance improvement per unit of parameter increase, it remains a cost-effective approach to map quantum state measurement probabilities to classical NN weights in the current stage.

\section{Conclusion and Future Work}
\label{sec:cfw}
In this work, we introduced significant enhancements to the QT framework by replacing the conventional MLP mapping model with a TN mapping model and implementing a distributed circuit ansatz to mitigate the challenges posed by large qubit requirements. These improvements enable more efficient parameter representation and reduce the computational overhead in both quantum hardware and classical simulation. Our results, validated on the CIFAR-10 classification task, show that the QT-TN model offers superior parameter efficiency compared to the QT-MLP baseline, requiring fewer parameters to reach comparable levels of accuracy.

A key observation from our findings is the different impact of increasing quantum versus classical parameters on model performance. While increasing quantum parameters, such as the QNN block repetition, provides greater accuracy gains per parameter unit, it also imposes significant demands on quantum coherence time and circuit simulation. In contrast, increasing classical parameters via the bond dimension of the tensor network provides a more scalable and practical solution in the noisy intermediate-scale quantum (NISQ) era, where quantum resources are limited and expensive to deploy.

From a broader perspective, this work demonstrates that hybrid quantum-classical frameworks like QT are highly adaptable. By introducing TNs, we open the possibility of leveraging classical techniques from quantum many-body physics to improve the scalability and performance of QML models. Additionally, our distributed circuit ansatz aligns with current trends toward modular and distributed quantum computation, positioning the QT framework as a forward-looking solution that can evolve alongside advancements in quantum hardware.

The significance of our contribution lies not only in improving the scalability of the QT framework but also in demonstrating a path forward for hybrid architectures that balance quantum and classical computation. By decoupling inference from quantum hardware and optimizing the training phase using TNs, we enable more flexible deployment of QML models in real-world applications. This adaptability makes the QT-TN framework particularly suitable for use cases in ML tasks that demand scalability, efficiency, and minimal reliance on quantum resources.

Our future work will focus on exploring more advanced tensor network architectures to further improve the efficiency and scalability of the QT framework. We aim to investigate alternative tensor decompositions, such as tree tensor networks (TTN) and projected entangled pair states (PEPS), which may offer better compression rates and accuracy trade-offs. Another area of interest is optimizing the distributed circuit ansatz to further reduce the qubit count without compromising model performance. Furthermore, testing the proposed framework on larger datasets and more complex tasks, such as reinforcement learning and natural language processing, will provide a broader assessment of its generalizability. Finally, integrating error mitigation techniques and optimizing the TN mapping model for real quantum hardware will be crucial in advancing the framework toward practical, real-world applications.


\bibliographystyle{IEEEtran}
\bibliography{bib/tools,bib/vqc,bib/qml_examples,bib/quantum_fl, bib/ml_examples, bib/hybrid_co_examples,bib/classical_fl, references}

\end{document}